\documentclass[twocolumn,aps,pra,showpacs,superscriptaddress]{revtex4-1}

\usepackage{amssymb}
\usepackage{mathrsfs}
\usepackage{lipsum}
\usepackage{graphicx}
\usepackage[caption=false]{subfig}
\usepackage{mathtools}
\usepackage{epstopdf}
\usepackage{xcolor}
\usepackage[colorlinks,urlcolor=blue]{hyperref}
\DeclareGraphicsExtensions{.pdf,.jpg,.png,.eps}
\usepackage[toc,page,title,titletoc,header]{appendix}
\begin{document}
	
	\title{Angular displacement estimation of Heisenberg scaling: Tunable squeezed Bell state via the enhancement of spin and orbital angular momenta}
	
	\author{Jian-Dong Zhang}
	\author{Zi-Jing Zhang}\email{Corresponding author: zhangzijing@hit.edu.cn}
	\author{Long-Zhu Cen}
	\author{Yuan Zhao}\email{Corresponding author: zhaoyuan@hit.edu.cn}
	\affiliation{Department of Physics, Harbin Institute of Technology, Harbin, 150001, China}

	\begin{abstract}
		We demonstrate an angular momentum-enhanced protocol that permits an angular displacement estimation by using tunable squeezed Bell state and parity detection. 
		We consider the resolution and the sensitivity, super-resolution is presented along with Heisenberg scaling sensitivity for arbitrary tunable factor, the tunable factor which can optimize the sensitivity is also discussed.
		Additionally, we analyze the advantages of using angular momentum via considering and comparing simulation results.
		Under the situation of the optimal tunable factor, the Heisenberg-limited sensitivity and $2\left(\ell+1\right)$-fold super-resolution peak with quantum number $\ell$ are achieved. 
	\end{abstract}
	
	
	\maketitle
	
	\section{INTRODUCTION}

	The field of quantum metrology exhibits excellent performance of utilizing quantum resources to boost the resolution and the sensitivity beyond what can be obtained with only classical resources. 
	In this regard, optical interferometers\cite{PhysRevA.33.4033} come across as ideal candidates and are of paramount meaning to the field of quantum metrology.
	In order to impel the sensitivity to reach the lower bound (quantum Cram\'er-Rao bound), innumerable researches have been carried out to explore improved methods.
	Generally, there are three classes of common technologies: non-classical input states, novel detection strategies, and amplitude or phase magnification via nonlinear processes.  
	
	A number of achievements have been achieved on the basis of the aforementioned three scenarios. 
	In the aspect of non-classical inputs, many a study shows the desirable results utilizing both Gaussian (single-mode \cite{PhysRevD.23.1693} and two-mode squeezed vacua \cite{PhysRevLett.104.103602}) and non-Gaussian states (N00N \cite{doi:10.1080/00107510802091298} and twin Fock states \cite{PhysRevA.68.023810}). 
	As to the novel detection strategies$\---$over the past few decades$\---$parity detection \cite{PhysRevLett.104.103602}, Z detection \cite{1367-2630-17-4-043030} and projective detection  \cite{PhysRevA.96.053846} have assisted plenty of systems in achieving ideal performances. 
	With regard to the nonlinear processes, SU(1,1) interferometer \cite{Chekhova:16} is the most typical application and shows the superior sensitivity than SU(2) case in some scenarios.

	Here we focus on estimating the angular displacement, which has become a valuable research in recent years, for it can be used to correct the measurement basis of two parties in quantum teleportation \cite{PhysRevA.83.053829,PhysRevA.96.053846, PhysRevLett.112.200401, d2013photonic}. 
	Under such a background, a lot of relevant schemes with using angular momentum are put forward, including the use of spin angular momentum (SAM) \cite{PhysRev.50.115} and orbital angular momentum (OAM) \cite{PhysRevA.45.8185}, related to the polarization and the spiral phase wavefront of light, respectively. 
	In turn, the sensitivities of Heisenberg scaling for both SU(1,1) and SU(2) interferometers are reported one after another.
	In this paper, we propose an estimation protocol toward angular displacement that possesses the sensitivity with Heisenberg scaling based upon tunable squeezed Bell (TSB) state and parity detection.
	Additionally, the combination of SAM and OAM is utilized to jointly enhance the estimation sensitivity.

	The remainder of the present paper is organized as follows: 
	in Sec. \ref{II}, we introduce the details of angular displacement estimation model, detection strategy, and the optimal polarized input state with considering polarized mode. 
	The input state, calculation process, and results are displayed in Sec. \ref{III}, additionally, the advantages of utilizing angular momentum is manifested. 
	In Sec. \ref{IV}, we discuss an optimal input, which comes from optimizing the tunable factor. 
	Finally, we conclude our work with a brief summary in Sec. \ref{V}.

	\begin{figure*}[t]
		\centering
		\includegraphics[width=\textwidth]{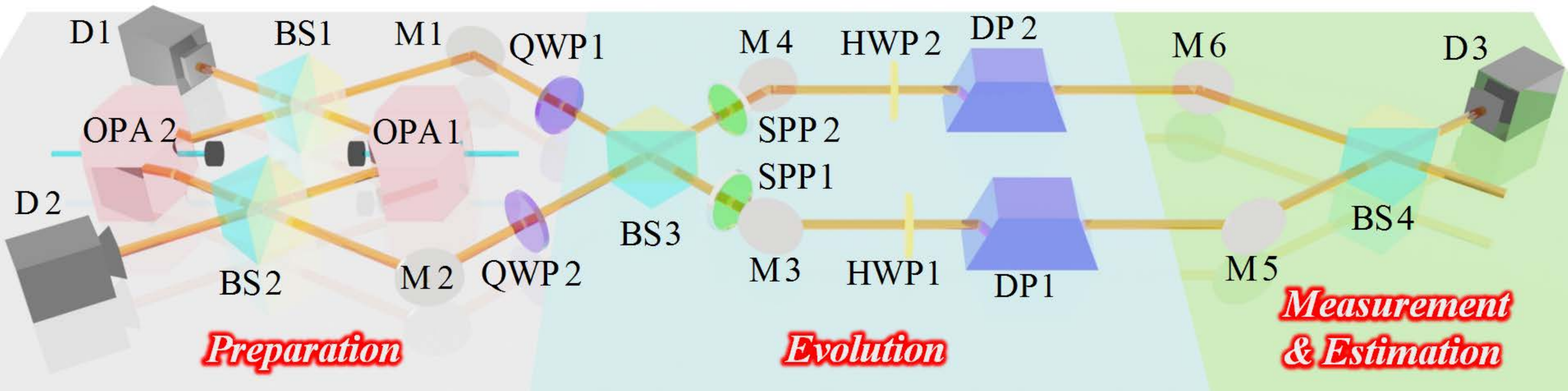}
		\caption{Schematic illustration of the estimation protocol for angular displacement. TSB state is produced in successful post-selection from the interaction of dual two-mode squeezed states which are from OPAs. The circular polarization character is equipped with QWPs and the SPPs give OAM to the TSB state. HWPs and DPs are used to modulate SAM and OAM, respectively, and single detector performs parity detection. OPA, optical parametric amplifier; BS, beam splitter; M, mirror; QWP, quarter wave plate; SPP, spiral phase plate; HWP, half wave plate; DP, Dove prism; D, detector.}
		\label{system}
	\end{figure*}
	
	\section{FUNDAMENTAL PRINCIPLE AND DEVICE}
	\label{II}
    
	Consider estimation protocol of the angular displacement whose input is generated by two independent Gaussian twin beams, as shown in Fig. \ref{system}, one of which plays the role of an ancillary two-mode state.
	The successful generation of the input state is triggered by two simultaneous detections for single photon realized by two detectors, and the details for the design of tunable factor can be found in Ref. \cite{PhysRevA.81.012333}.
	Then the input is transfered to a specific polarized state and enters the interferometer, subsequently, the state carries OAM with the help of SPPs.
    Two sets of rotating devices (two DPs and two HWPs) are inserted into two paths, they not only introduce a phase shift in two paths but also rotate the polarization of the quantum state.
	Moreover, the DP and the HWP are rotated the identical angle to simulate the rotation of SAM and that of OAM around the optical axis.
	After such evolution as above, finally, the output detection is performed.

	Parity detection is the adopted strategy in our protocol.
	It was first proposed when Bollinger and colleagues studied trapped ions \cite{PhysRevA.54.R4649}.
	As a binaryzation method, the interest of parity detection lies in distinguishing the parity of photon number, instead of exact number of photons, at either of the two output ports.
	+1 is recorded for appearing even photons and --1 for odd ones. 
	In views of the above depiction, in turn, the parity operator for path $A$ is given by ${\hat \Pi _A} = \exp \left( {i\pi {{\hat n}_A}} \right)$.
	 Furthermore, it has been proved that parity detection is the optimum strategy for a great deal of quantum states in optical interferometers, especially for the path symmetry states \cite{PhysRevA.87.043833}.

	Because of the introduction of polarized information, initially, we discuss the optimal polarized mode for such an estimation protocol and segment the input state according to the orthogonally polarized modes in paths $A$ and $B$.
	For the sake of simplicity, hereon we only take the coherent state and parity detection as an example. 
	Within foregoing description we can provide the input with the following representation,
	\begin{equation}
	\left| {{\psi _\textrm{0}}} \right\rangle  = {\left( {{{\left| \alpha  \right\rangle }_H}{{\left| {\beta {e^{i\varphi }}} \right\rangle }_V}} \right)_A}{\left( {{{\left| 0 \right\rangle }_H}{{\left| 0 \right\rangle }_V}} \right)_B}.
	\end{equation}
	Where $\alpha$ and $\beta$ stand for the complex amplitude of coherent states in horizontal mode ($H$) and vertical one ($V$), respectively.
	The parameter $\varphi$ indicates the phase difference between the two polarized modes. 
	Thus an arbitrary pure polarized state can be denoted by the state ${{{\left| \alpha  \right\rangle }_H}{{\left| {\beta {e^{i\varphi }}} \right\rangle }_V}}$.
	After the first beam splitter, the state arrives at 
	$\left| {{\psi _1}} \right\rangle  = {\left( {{{\left| {{\alpha  \mathord{\left/
								{\vphantom {\alpha  {\sqrt 2 }}} \right.
								\kern-\nulldelimiterspace} {\sqrt 2 }}} \right\rangle }_H}{{\left| {{{\beta {e^{i\varphi }}} \mathord{\left/
								{\vphantom {{\beta {e^{i\varphi }}} {\sqrt 2 }}} \right.
								\kern-\nulldelimiterspace} {\sqrt 2 }}} \right\rangle }_V}} \right)_A}{\left( {{{\left| {i{\alpha  \mathord{\left/
								{\vphantom {\alpha  {\sqrt 2 }}} \right.
								\kern-\nulldelimiterspace} {\sqrt 2 }}} \right\rangle }_H}{{\left| {{{i\beta {e^{i\varphi }}} \mathord{\left/
								{\vphantom {{i\beta {e^{i\varphi }}} {\sqrt 2 }}} \right.
								\kern-\nulldelimiterspace} {\sqrt 2 }}} \right\rangle }_V}} \right)_B}$.
	Then it experiences two sets of rotating devices which are orientated with an angular displacement difference of $\phi$,  
	To such a angular displacement difference there corresponds to a relative phase difference of $2\ell\phi$ between the two paths in the interferometer \cite{lavery2011robust, zhang2014mimicking}.
	Meanwhile, a polarized rotation angle of $2\phi$ is also generated in two modes. 
	Hence the state can be written as
	\begin{widetext}
		\begin{eqnarray}
		\nonumber \left| {{\psi _2}} \right\rangle  =&& {\left( {{{\left| {\frac{{{e^{i2\ell\phi }}}}{{\sqrt 2 }}\left[ {\alpha \cos \left( {2\phi } \right) - \beta {e^{i\varphi }}\sin \left( {2\phi } \right)} \right]} \right\rangle }_H}{{\left| {\frac{{{e^{i2\ell\phi }}}}{{\sqrt 2 }}\left[ {\alpha \sin \left( {2\phi } \right) + \beta {e^{i\varphi }}\cos \left( {2\phi } \right)} \right]} \right\rangle }_V}} \right)_A} \\
		&&\otimes  {\left( {{{\left| {\frac{{i\alpha }}{{\sqrt 2 }}} \right\rangle }_H}{{\left| {\frac{{i\beta }}{{\sqrt 2 }}{e^{i\varphi }}} \right\rangle }_V}} \right)_B}.
		\end{eqnarray}
		Then the state passes through the second beam splitter and the output reads 
		\begin{equation}
		{\left| {{\psi _\textrm{3}}} \right\rangle _A} = {\left| {\frac{{{e^{i2\ell\phi }}}}{2}\left[ {\alpha \cos \left( {2\phi } \right) - \beta {e^{i\varphi }}\sin \left( {2\phi } \right)} \right] - \frac{\alpha }{2}} \right\rangle _H}{\left| {\frac{{{e^{i2\ell\phi }}}}{2}\left[ {\alpha \sin \left( {2\phi } \right) + \beta {e^{i\varphi }}\cos \left( {2\phi } \right)} \right] - \frac{\beta }{2}{e^{i\varphi }}} \right\rangle _V}.
		\label{e3}
		\end{equation}
		
	\end{widetext}
	
	Note that the state in Eq. (\ref{e3}) is a reduced output for path $A$ since parity detection only needs to monitor a single output port.
    Consider two pure polarized scenarios: linear polarization (${\left| \alpha  \right|^2} =  N_\textrm{C} $, $\left|\beta \right|= 0$ and $\varphi  =  0$) and circular polarization (${\left| \alpha  \right|^2} = {\left| \beta  \right|^2} = {N_\textrm{C}/2}
	$ and $\varphi  =  - \pi /2$).
	The corresponding expectation values of parity detection are given by
	\begin{eqnarray}
	{\left\langle {{{\hat \Pi }_A}} \right\rangle _\textrm{LP}} &&= \exp \left\{ { - N_{\rm C}\left[ {1 - \cos \left( {2\phi } \right)\cos \left( {2\ell\phi } \right)} \right]} \right\}, \\
	{\left\langle {{{\hat \Pi }_A}} \right\rangle _\textrm{CP}} &&= \exp \left\{ { - 2N_{\rm C}{{\sin }^2}\left[ {\left( {\ell + 1} \right)\phi } \right]} \right\}.
	\label{ccc}
	\end{eqnarray}

	\begin{figure}[htbp]
		\centering
		\includegraphics[width=8cm]{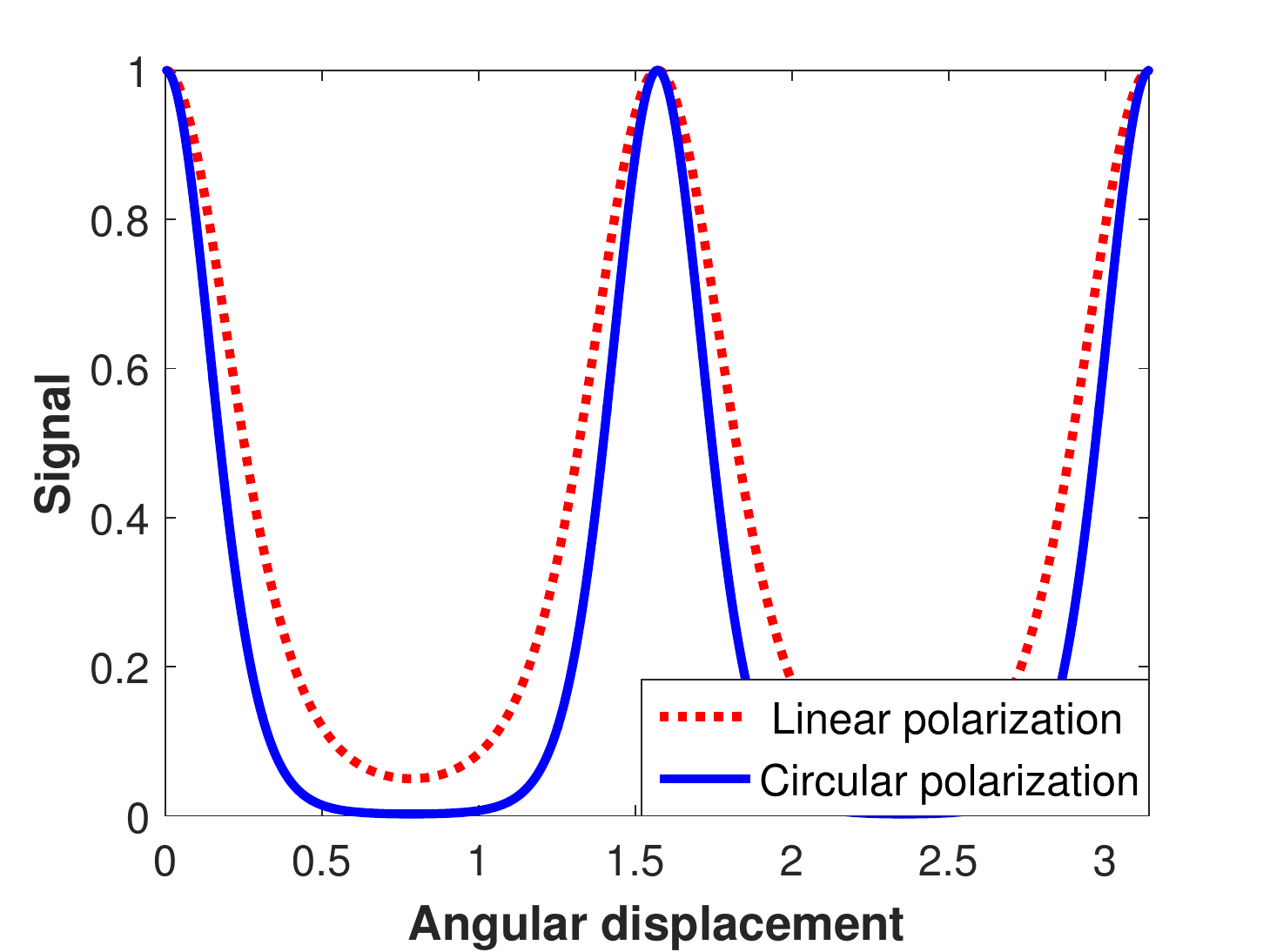}
		\caption{The signals of circularly polarized state and linearly polarized one with parity detection as functions of angular displacement in the case of $\ell=1$ and $N_\textrm{C}=3$. }
		\label{cp}
	\end{figure}
	
   As for the case of ${\left| \alpha  \right|^2} = {\left| \beta  \right|^2} = {N_\textrm{C}/2}$ and $\varphi  =  \pi /2$, we only need to replace $\ell$ with $-\ell$. 
	This operation is not a tricky problem since this can be matched at the input state.

	We plot Fig. \ref{cp} with the parameters $\ell=1$ and $N_\textrm{C}=3$ to intuitively observe the impact of two polarized states on the detection result.
	From the figure one can see that both the signal visibility and the full wave at half maximum (FWHM) of linearly polarized input are inferior to these of circularly polarized one, i.e., circularly polarized state has a better visibility and super-resolution character
	Based upon this basic, in the following sections, we adopt circularly polarization in protocol.  
	In addition, the term $(\ell+1)\phi$ in Eq. (\ref{ccc}) shows that SAM has the same physical effect as OAM.
	This also reveals the fact that circularly polarized photons carry spin angular momentum and are eigenstates of rotating operation, i.e., the rotation of the linearly polarized mode is equivalent to introducing equivalent and reverse phase shift in two circularly polarized modes.

	\section{THEORETICAL ANALYSIS}
	\label{III}
	Squeezed states have been studied throughly in the field of parameter estimation. 
	Exotic results, both theoretical and experimental, are popping out from time to time. 
	The physical mechanism toward the enhancement sensitivity based on the single-mode squeezed state is to reduce the noise fluctuation.
	As for two-mode squeezed state, the strong correlation between the two modes is the primary cause for improving sensitivity.
	Here, we consider TSB state, which is originally introduced in Refs. \cite{PhysRevA.76.022301,PhysRevA.81.012333} and is of the form as follows (see Appendix for derivation)
	\begin{widetext}
		\begin{eqnarray}
		\nonumber{\left| \psi  \right\rangle _\textrm{TSB}} =&& {\hat S_{12}}\{ \cos\delta \left| {0,0} \right\rangle  + \sin \delta \left| {1,1} \right\rangle \} \\
		\nonumber=&& \sum\limits_{n = 0}^\infty  {\{ [\frac{{\cos \delta }}{{\cosh r}}} {(-\tanh r)^n}] + \sin \delta {(-\tanh r)^{n - 1}}[n{C_{1,1}} + \tanh r(n - 1){C_{0,0}}]\} \left| {n,n} \right\rangle \\
		=&& \sum\limits_{n = 0}^\infty  {G(n} )\left| {n,n} \right\rangle.
		\label{TSB}
		\end{eqnarray}
	\end{widetext}
	Where the parameters are given by
	\begin{eqnarray}
	{C_{0,0}} &&= \left\langle {0,0} \right|S\left( \xi  \right)\left| {1,1} \right\rangle  = \frac{{\tanh r}}{{\cosh r}}, 
	\label{00}\\
	{C_{1,1}} &&= \left\langle {1,1} \right|S\left( \xi  \right)\left| {1,1} \right\rangle  = \frac{{1 - {{\sinh }^2}r}}{{{{\cosh }^3}r}}.
	\label{11}
	\end{eqnarray}
	While $S\left( \xi  \right)$ is the two-mode squeezing operator, $\delta$ is a free tunable factor allowing for choice by adjusting experimental parameters, and $r$ is the squeezing factor. 
	With some suitable values of $\delta$, the TSB state can be used to simulate many two-mode states, e.g., photon-added/subtracted squeezed state, squeezed vacuum state and squeezed number state. 
	
	In views of the parity detection mentioned earlier and the expanded form of the TSB state in twin Fock representation, we can calculate the expectation value of parity operator for port $A$,
	\begin{eqnarray}
	\nonumber\left\langle {{\hat \Pi _A}} \right\rangle  =&& \left\langle {{\psi _\textrm{out}}} \right|{\hat \Pi _A}\left| {{\psi _\textrm{out}}} \right\rangle \\
	\nonumber=&& {}_\textrm{TSB}\left\langle \psi  \right|{\hat U^\dag }{\hat \Pi _A}\hat U{\left| \psi  \right\rangle _\textrm{TSB}}\\
	\nonumber=&& \sum\limits_{n = 0}^\infty  {{G^\dag }(n)G(n} )\left\langle {n,n} \right|{\hat U^\dag }{\hat \Pi _A}\hat U\left| {n,n} \right\rangle \\
	=&& \sum\limits_{n = 0}^\infty  \left|{G(n)}\right| ^2{P_n}\left\{-\cos[4(\ell+1)\phi ]\right\}.
	\label{Legendre}
	\end{eqnarray}
	Where ${P_n}\left({s}\right)$ is $n$-order Legendre polynomials. With the help of error propagation, we can obtain the sensitivity
	\begin{equation}
	\Delta \phi  = \frac{{\sqrt {1 - {{\left\langle {{{\hat \Pi }_A}} \right\rangle }^2}} }}{{\left| {{{\partial \left\langle {{{\hat \Pi }_A}} \right\rangle } \mathord{\left/
						{\vphantom {{\partial \left\langle {{{\hat \Pi }_A}} \right\rangle } {\partial \phi }}} \right.
						\kern-\nulldelimiterspace} {\partial \phi }}} \right|}}.
	\label{sen}
	\end{equation}
	
\begin{figure}[htbp]
		\centering
		\includegraphics[width=8cm]{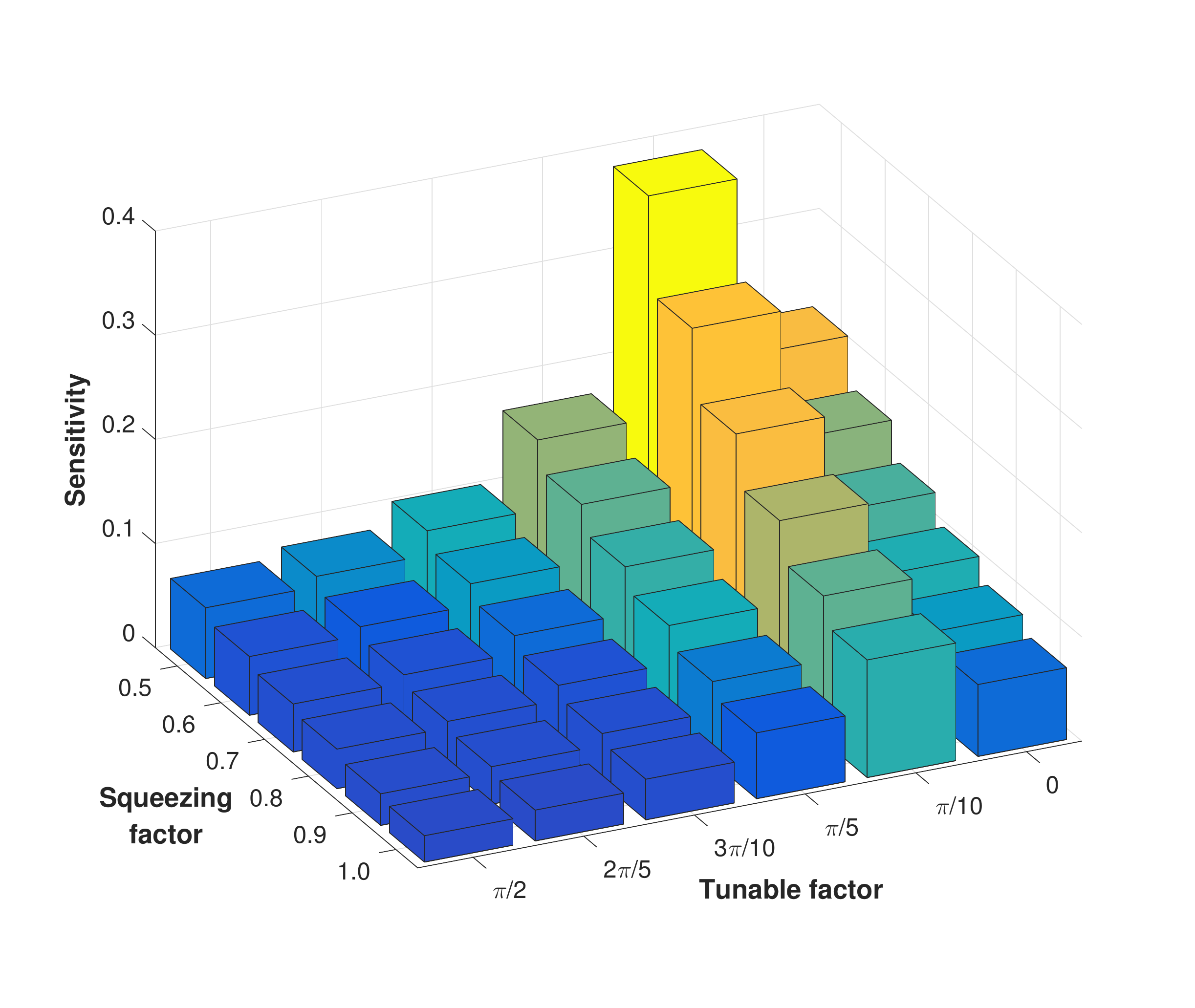}
		\caption{The sensitivity of TSB state as a function of both $r$ and $\delta$ in the case of $\ell=1$. The value ranges of  $r$ and  $\delta$ are between 0.5 to 1 and between 0 to ${\pi}/2$, respectively.}
		\label{whole}
	\end{figure}

	Since this sensitivity has no intuitive expression, we plot Fig. \ref{whole} by numerical method to clearly observe the results of  angular displacement estimation based on the TSB state.
	Figure \ref{whole} manifests the variation of optimal sensitivities with different $r$ and $\delta$.
	The optimal sensitivity raises with the increase of $r$, however, one interesting phenomenon is that the relationship between sensitivity and $\delta$ is not monotonous,  an inflection point appears at the position of $\delta=\pi/10$.

\begin{figure}[htbp]
		\centering
		\includegraphics[width=4cm]{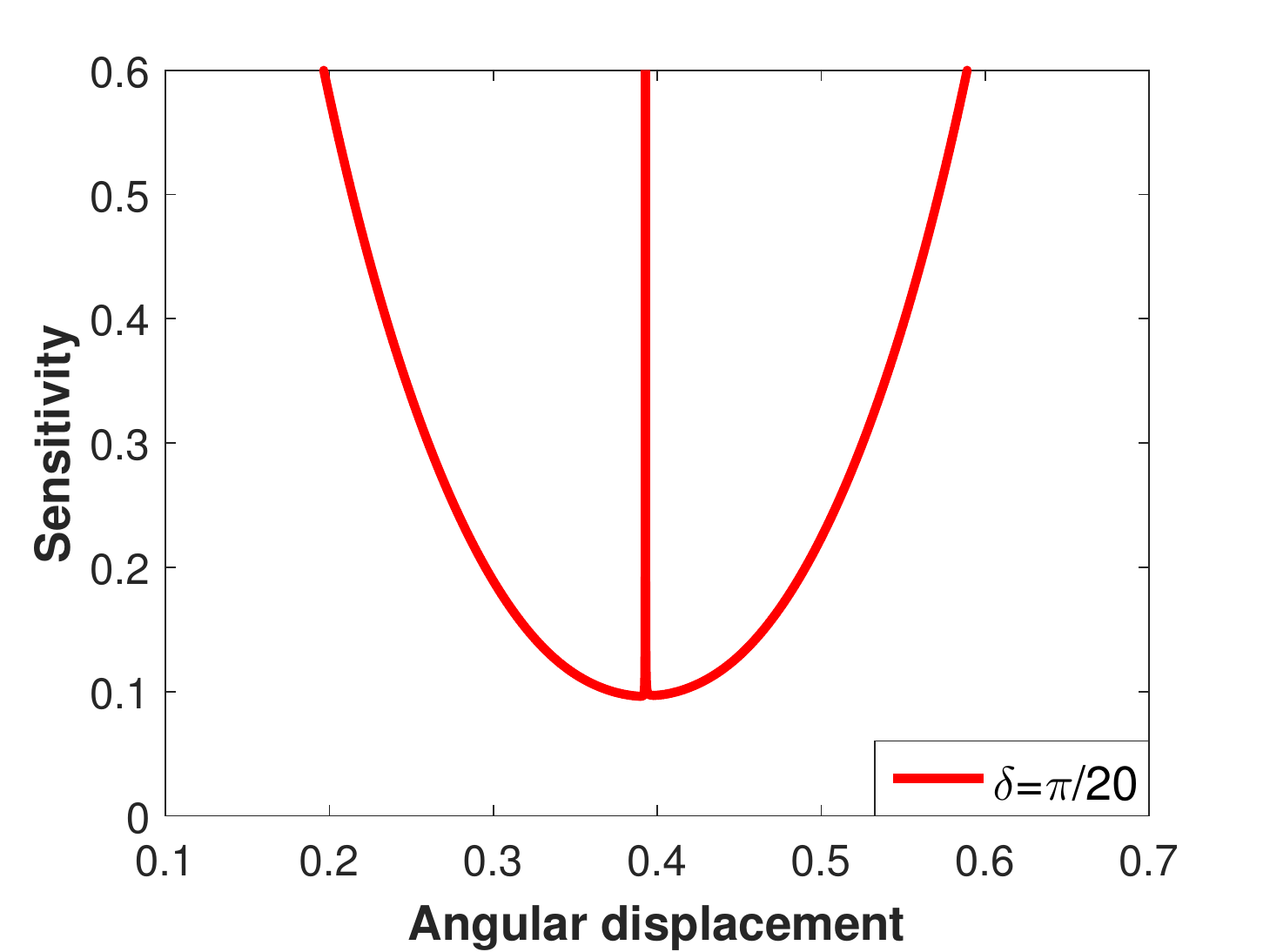}
		\includegraphics[width=4cm]{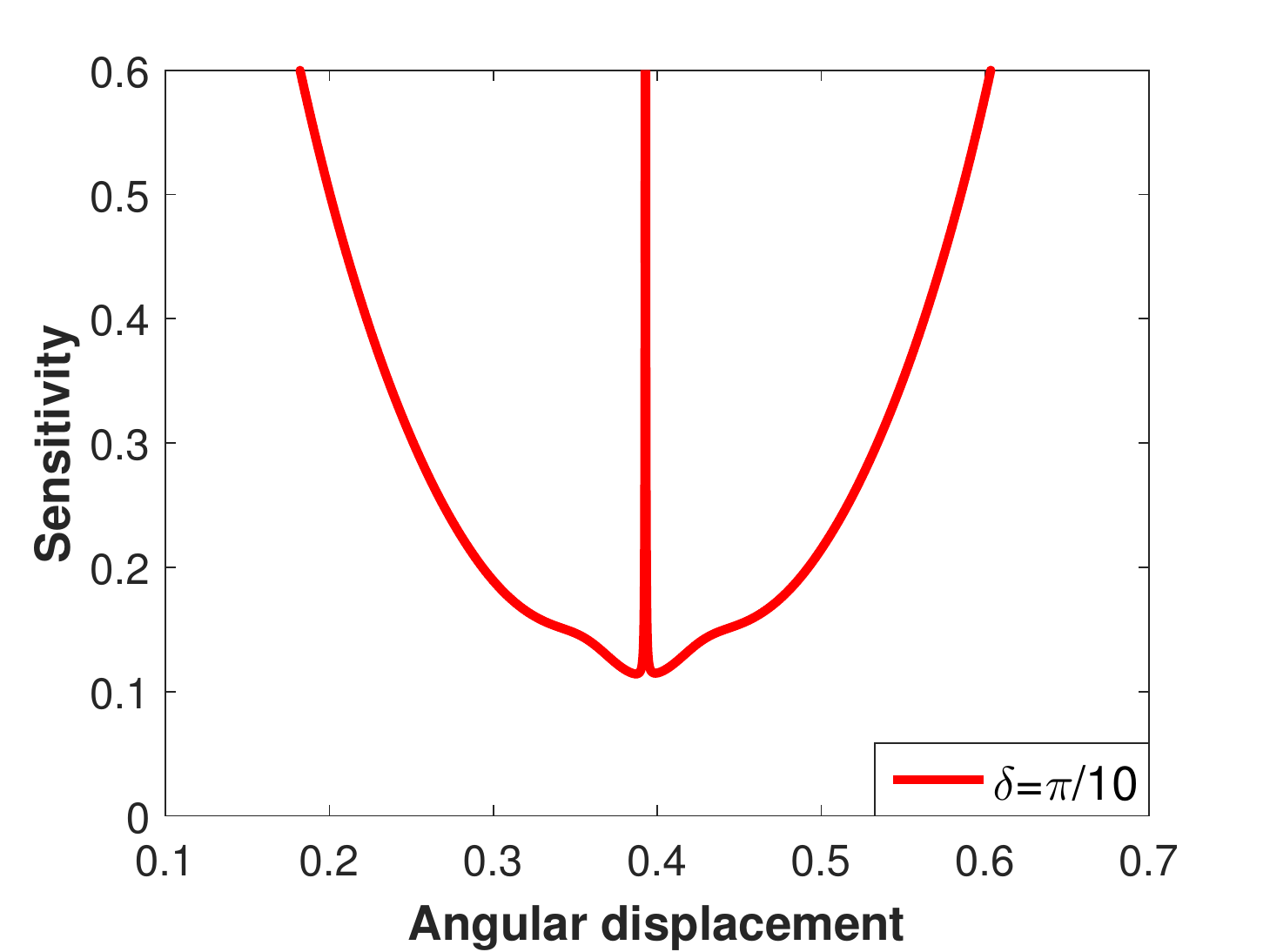}
		\includegraphics[width=4cm]{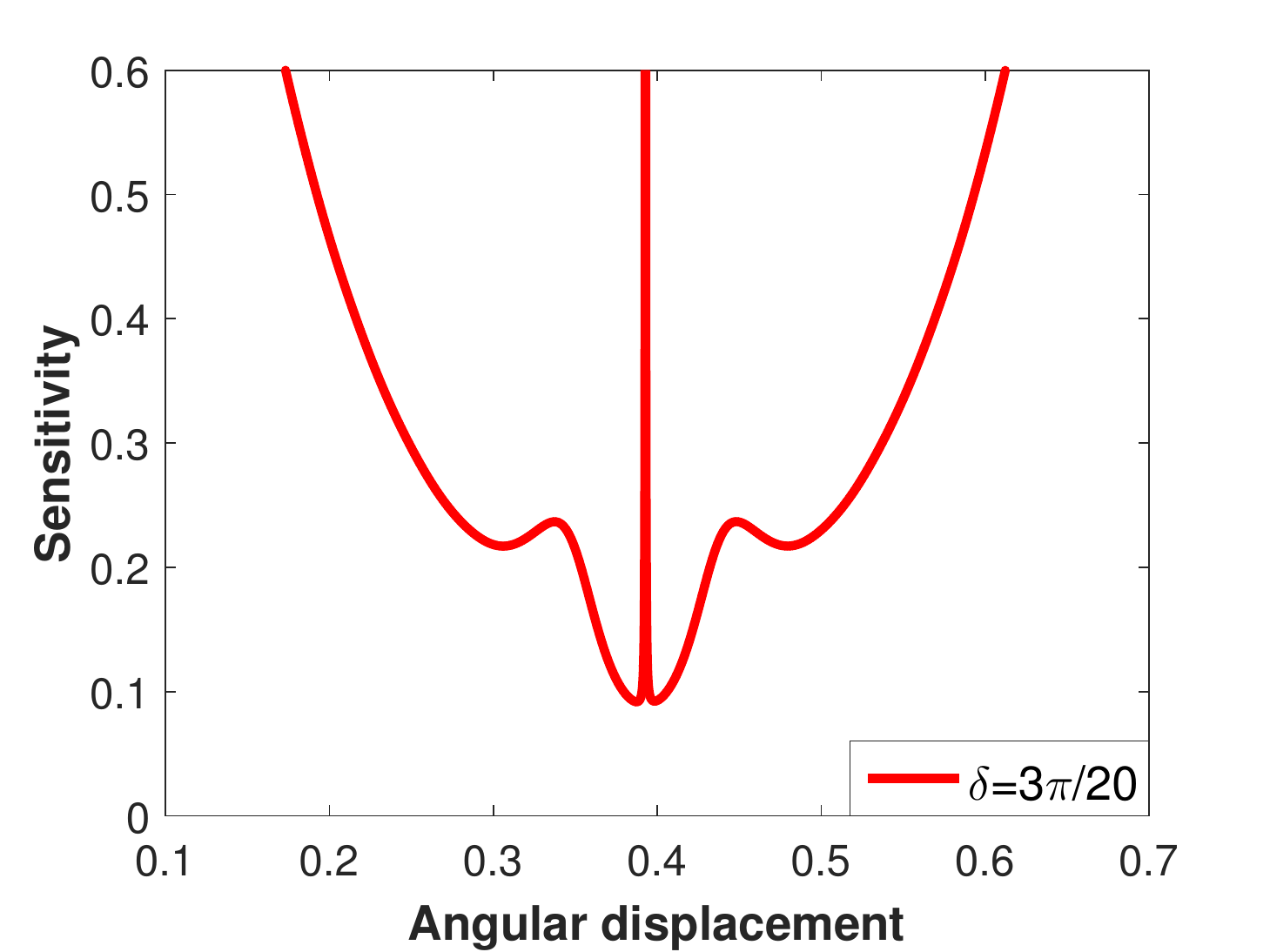}
		\includegraphics[width=4cm]{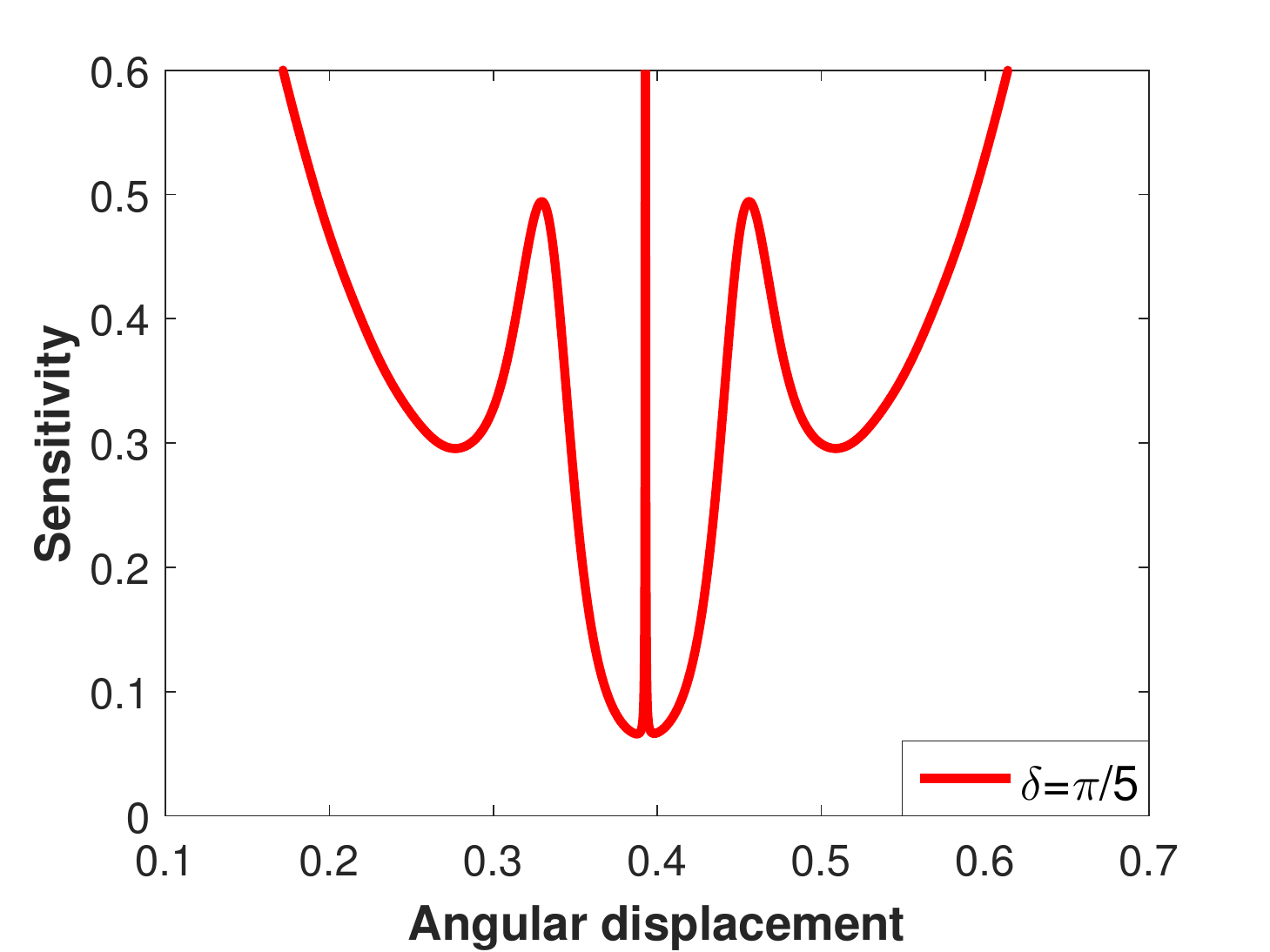}
		\includegraphics[width=4cm]{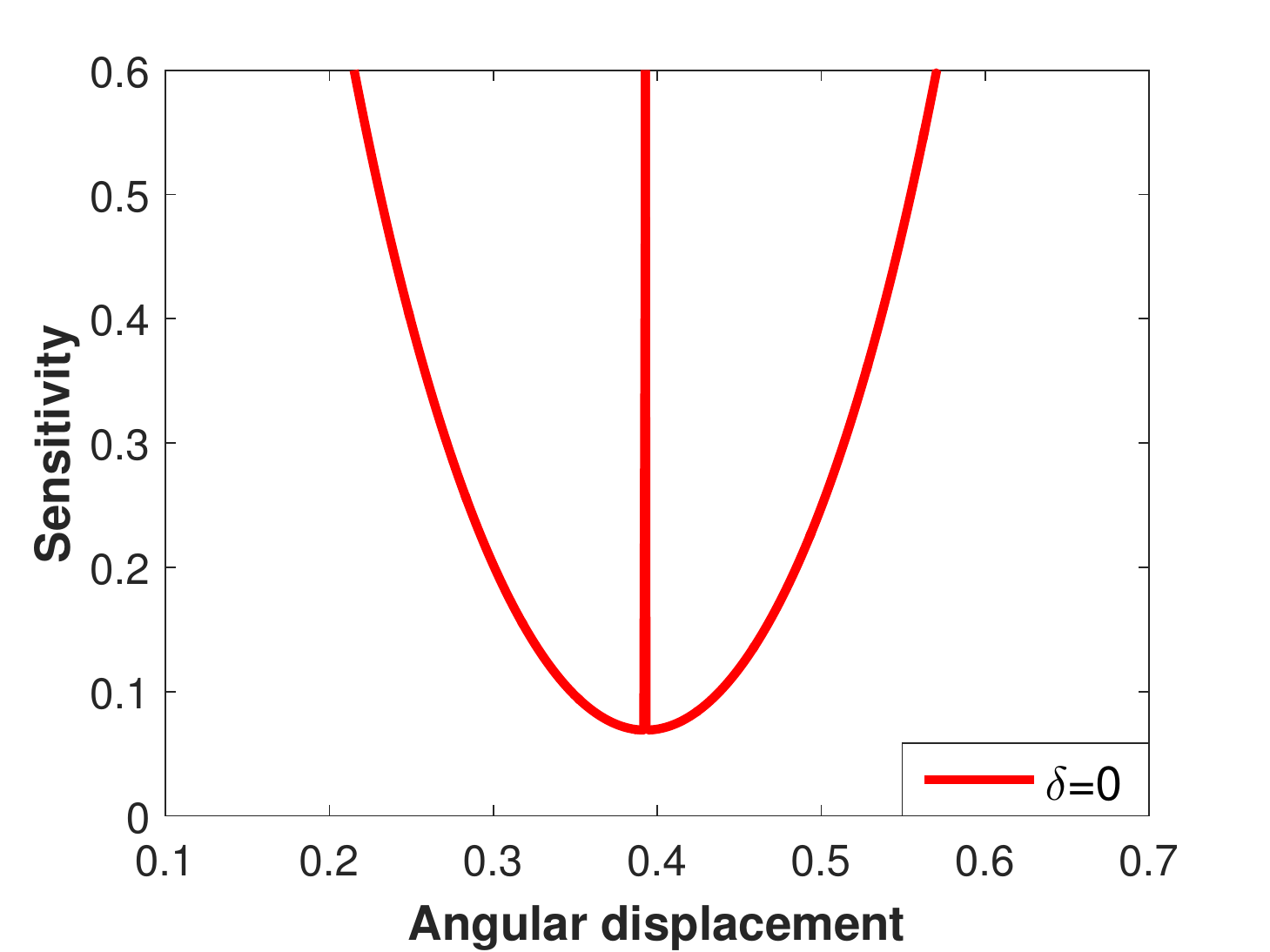}
		\includegraphics[width=4cm]{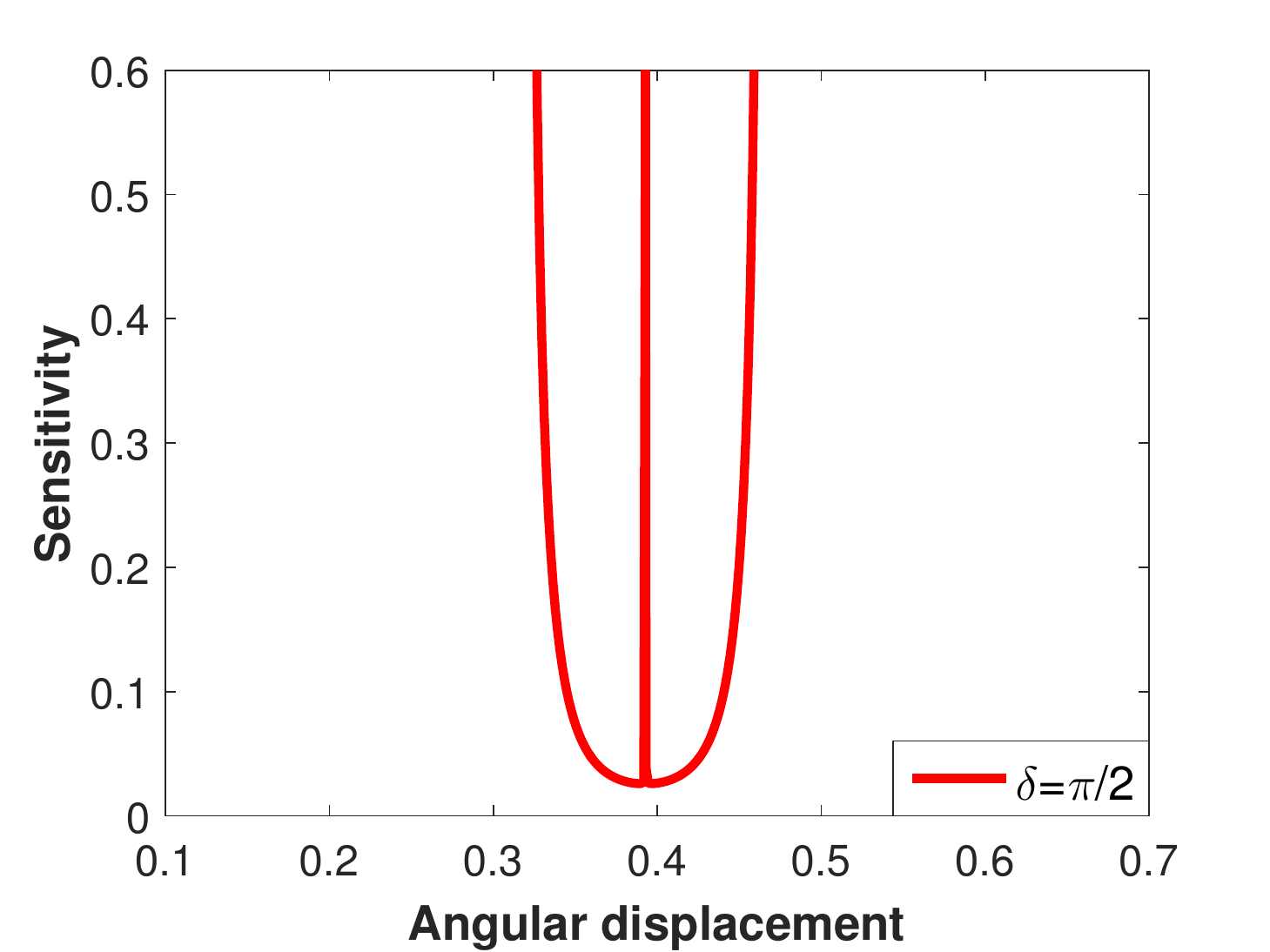}
		\caption{The sensitivity of TSB state as a function of angular displacement with $\ell=1$, $r=1$ and different $\delta$. The last two subfigures are two extreme cases: two-mode squeezed vacuum state ($\delta=0$) and two-mode squeezed number state ($\delta=\pi/2$).}
		\label{fff}
	\end{figure}

	To further explore the reason behind this situation, we plot Fig. \ref{fff} with different $\delta$ near the $\pi/10$.
	The first subgraph points out that the sensitivity at this time is mainly led by the two-mode squeezed vacuum state.
	With the further increase of $\delta$, the weight of two-mode squeezed number state is rising and that of squeezed vacuum one is declining, as illustrated in second subfigure.
	Note that, for the same $r$, the sensitivity of squeezed number state is better than that of squeezed vacuum.
	As for the third and the fourth cases, the weight of squeezed number state is dominant, so that the optimal sensitivity continues to rise for large $\delta$.
	The last two subfigures$\---$two-mode squeezed vacuum ($\delta=0$) and two-mode squeezed number states ($\delta=\pi/2$)$\---$serve as the control group.
	On the whole, the fundamental cause for the appearance of inflection point stems from the weight game.

\begin{figure}[htbp]
		\centering
		\includegraphics[width=8cm]{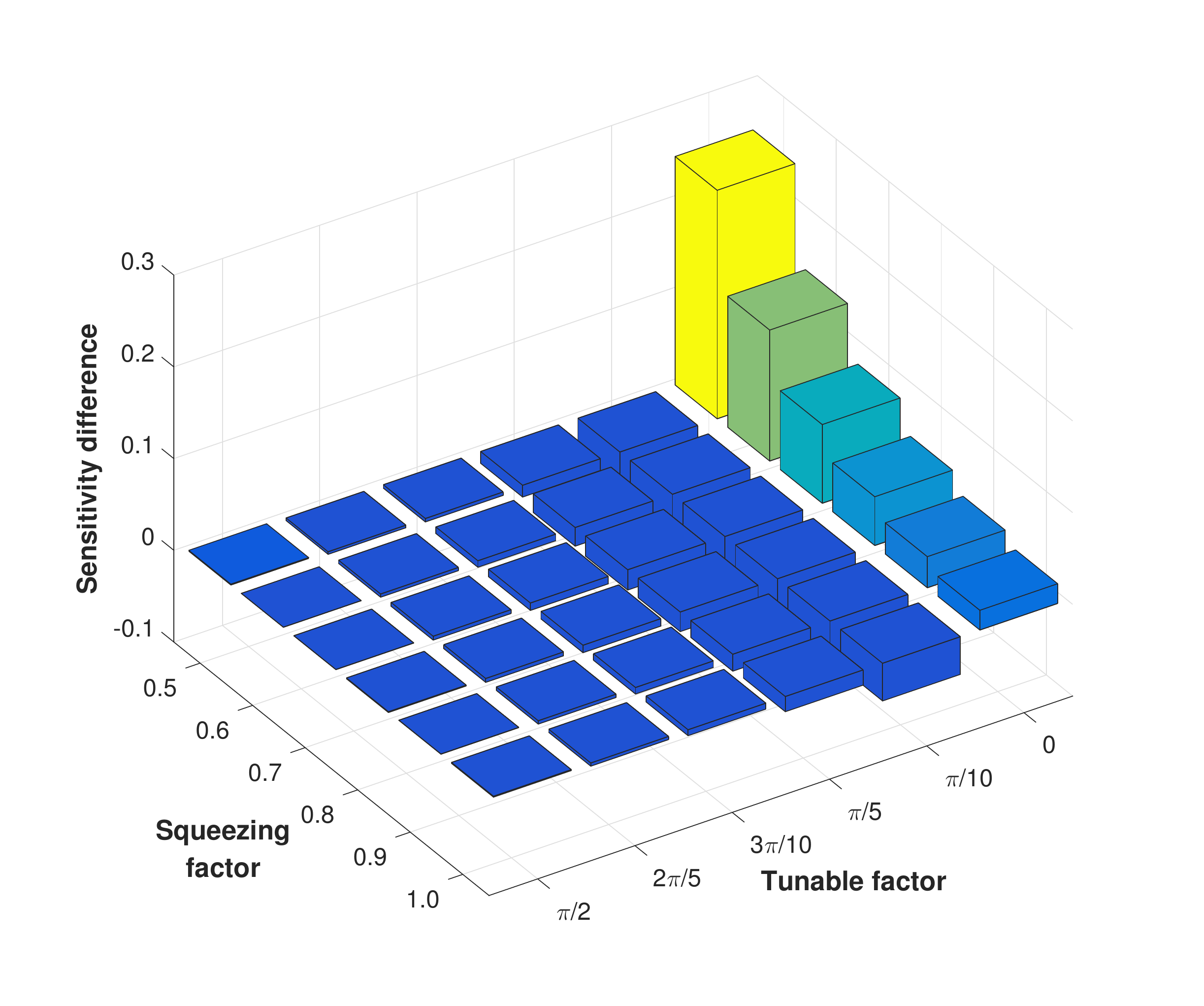}
		\caption{The sensitivity difference between TSB state with parity detection and Heisenberg limit, $\Delta\phi_\textrm{HL}-\Delta\phi$,  as a function of both $r$ and $\delta$ in the case of $\ell=1$. The value ranges of  $r$ and $\delta$ are between 0.5 to 1 and between 0 to ${\pi}/2$, respectively.}
		\label{HL}
	\end{figure}

	For the sake of objectively evaluating the sensitivity of our protocol, we compare the sensitivity of parity detection and the Heisenberg limit, as shown in Fig. \ref{HL}.
	The definition of sensitivity difference in the figure is $\Delta\phi_\textrm{HL}-\Delta\phi$, in turn, positive value represents sub-Heisenberg-limited sensitivity.
	One obvious conclusion is that the two-mode squeezed vacuum ($\delta=0$) can break the Heisenberg limit.
	For other cases, with the increase of $\delta$, the optimal sensitivity is approximated to Heisenberg limit, especially for $\delta={\pi/2}$ (saturation in the Heisenberg limit). 
	Combined with Figs. \ref{whole} and \ref{HL},  the case of $\delta={\pi/2}$ is the optimal sensitivity although it is not as obvious as the case of $\delta=0$ to beat the Heisenberg limit.
	
\begin{figure}[htbp]
		\centering
		\includegraphics[width=8cm]{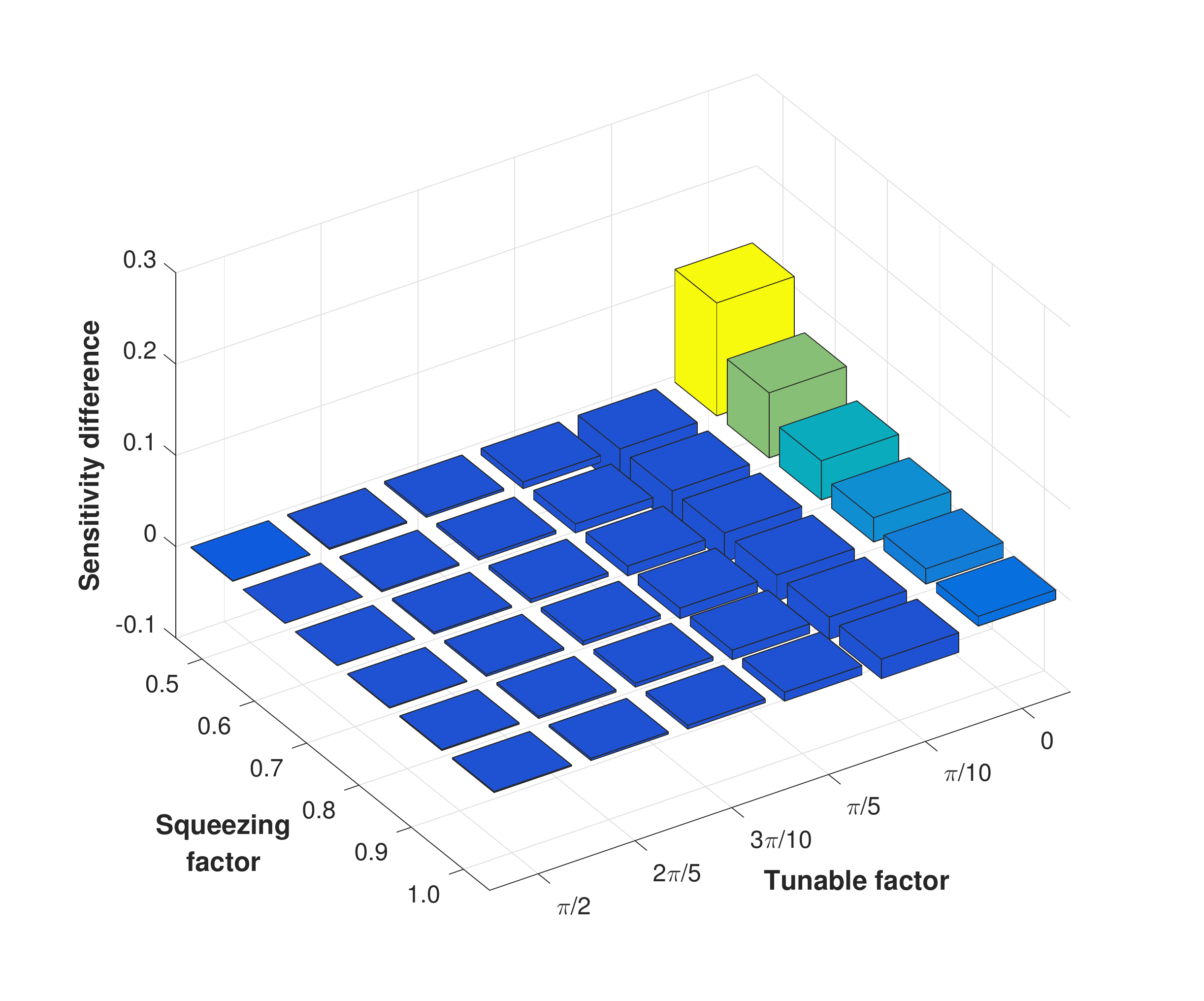}
		\caption{Same as Fig. \ref{HL}, but for $\ell=3$.}
		\label{HL2}
	\end{figure}

	As the final content of this section, we explore the impacts of angular momentum on resolution and sensitivity.
	Apropos of resolution, the item ${P_n}\left[\cos[4(\ell+1)\phi ]\right]$ in Eq. (\ref{Legendre}) indicates that the signal has a multi-fold super-resolution peak.
	For sensitivity, we plot Fig. \ref{HL2} via choosing the same parameters with Fig. \ref{HL} except for $\ell=3$.
	One sees that the variation trends in two figures are identical, however, there is a diminution for sensitivity difference, i.e., by raising the quantum number, the sensitivity of the system can asymptotically reach the Heisenberg limit.
	
	Here we give a brief summary for the performance improvement of our system. 
	There are three main advantages for using spin and orbital angular momenta to boost estimation. 
	Initially, since the participation of OAM and SAM, single-fold super-resolution under the situation of non-OAM protocol is extended to the $2\left(\ell+1\right)$-fold super-resolution in our protocol, which can be used in the field of lithography \cite{PhysRevLett.85.2733}.
	Furthermore, the scanning range in the practical measurement is significantly shortened, for the multi-fold signal peak brings about multiple optimal positions for sensitivity.
	Eventually, the sensitivity is boosted by a factor of $2\left(\ell+1\right)$ with the assist of two kinds of angular momenta, and this effect is equivalent to repeating $4\left(\ell+1\right)^2$ times in the protocols without using  angular momentum.
	The difference between the sensitivity and the Heisenberg limit is also compressed,  an asymptotically optimal effect, with increasing quantum number of OAM.

	\section{OPTIMAL SENSITIVITY: TWO-MODE SQUEEZED NUMBER STATE}
	\label{IV}
	It can be seen that the optimal sensitivity appears at $\delta={\pi/2}$ from the discussion in the previous section.
	This implies that the optimal state for sensitivity is two-mode squeezed number state, which can be expanded into the form as below,
	\begin{equation}
	{\left| \psi  \right\rangle _\textrm{SN} } = \sum\limits_{n = 0}^\infty  {\frac{{{{\left( { - \tanh r} \right)}^{n - 1}}}}{{{{\cosh }^3}r}}\left( {n - {{\sinh }^2}r} \right)\left| {n,n} \right\rangle }. 
	\end{equation}
	The mean photon number is offered by $N_\textrm{SN} = \textrm{Tr}\left[ {{\rho_\textrm{SN}}\left( {{{\hat a}^\dag }\hat a + {{\hat b}^\dag }\hat b} \right)} \right] = 6{\sinh ^2}r + 2$.
	Unlike the other extreme case of TSB state (two-mode squeezed vacuum), TMSN state is a non-Gaussian state.
	
\begin{figure}[htbp]
		\centering
		\includegraphics[width=8cm]{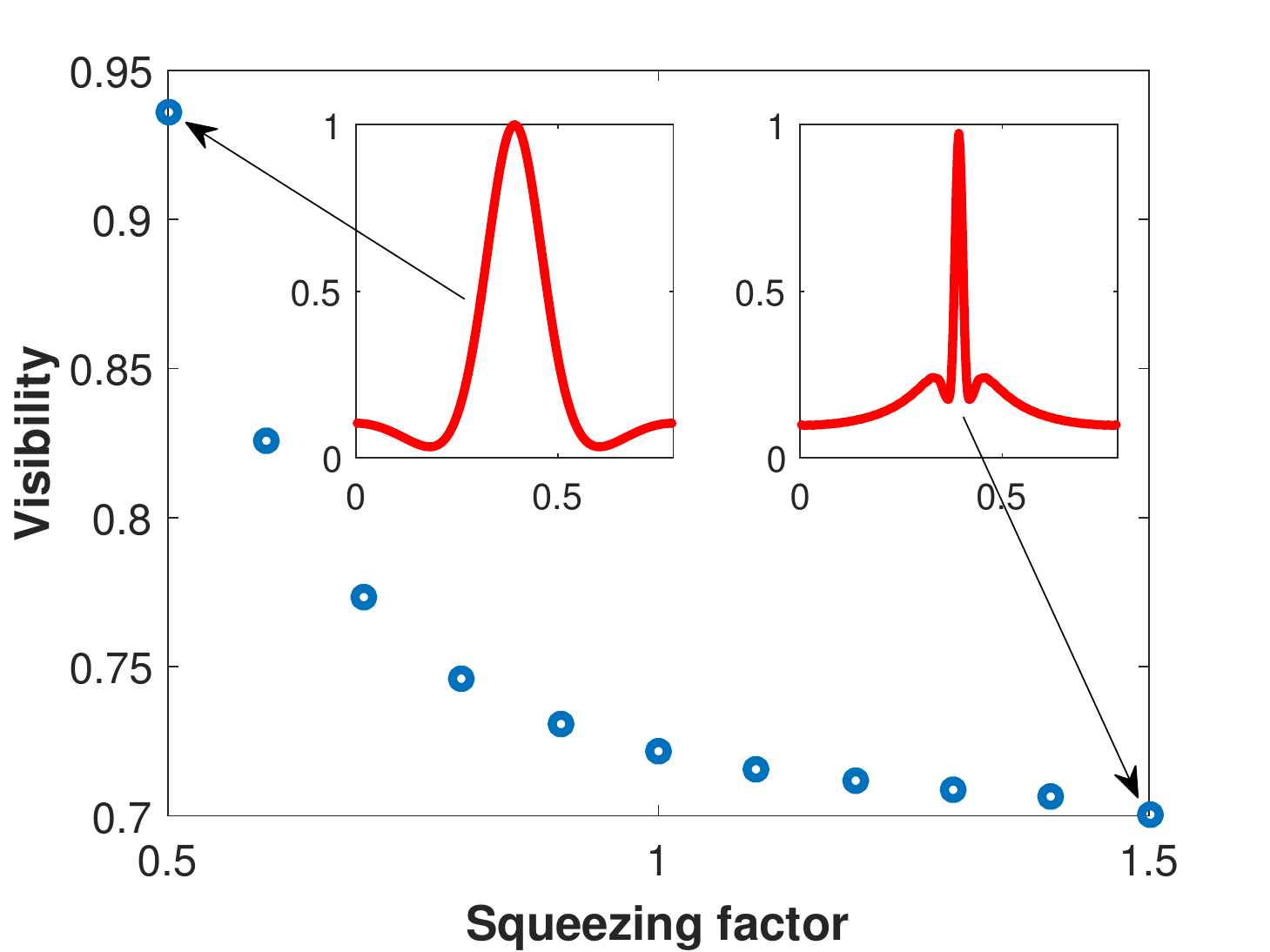}
		\caption{The visibility of TMSN state as a function of $r$ in the case of $\ell=1$. The value range of $r$ is between 0.5 to 1.5. The left subgraph and the right one correspond to the expectation values of $r=0.5$ and $r=1.5$, respectively.}
		\label{visibility}
	\end{figure}

	By utilizing the numerical method we plot the signal visibility of the TMSN state, as displayed in Fig. \ref{visibility}.
	From the figure we can see that  the visibility of signal decreases and the variation tends to gradually steady with the increase of $r$. 
	Further increasing $r$ has no obvious impact on the visibility.
	In addition, the FWHM of signal gets narrower with raising $r$, i.e., the super-resolution character becomes more significant when to raise $r$.
	
\begin{figure}[htbp]
		\centering
		\includegraphics[width=8cm]{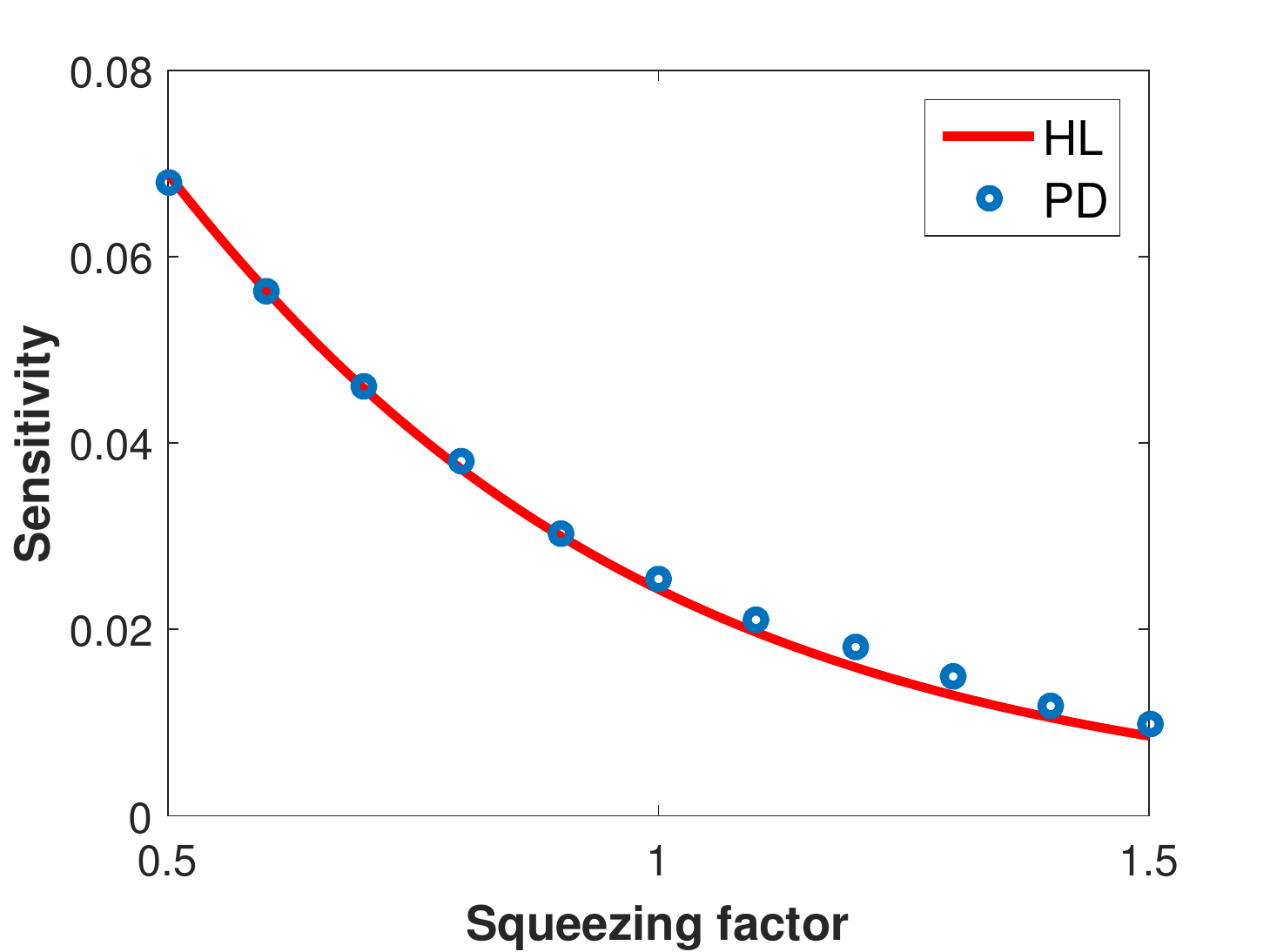}
		\caption{The sensitivity of TMSN state as a function of $r$ in the case of $\ell=1$. The value range of $r$ is between 0.5 to 1.5. HL, Heisenberg limit; PD, parity detection.}
		\label{sensitivity}
	\end{figure}

	We also study the sensitivity of the TMSN state, and the result compared with the Heisenberg limit is shown in Fig. \ref{sensitivity}.
	The simulation illuminates that the optimal sensitivity of parity detection is basically the same as the Heisenberg limit.
	It also need to point out that the mean photon number of the TMSN state is approximately 3 times as large as that of the two-mode squeezed vacuum, i.e., a 3 factor enhancement increase towards sensitivity compared to the squeezed vacuum. 
	
	As an extreme case of the TSB state, the conclusion about angular momentum advantages described in the Sec. \ref{III} is fully applicable to the TMSN state.

	\section{CONCLUSION}
	\label{V}
	To summarize, we report on a protocol that utilizes tunable squeezed Bell state containing spin and orbital angular momenta to perform angular displacement estimation.
	The results illustrate that our protocol achieves $2\left(\ell+1\right)$-fold super-resolution peaks and Heisenberg scaling sensitivity, and we study the variation in tunable squeezed Bell state with squeezing and tunable factors.
	We also discuss the advantages of utilizing angular momentum, including expansion of multi-fold super-resolution peaks along with shortening of the scanning range, and an asymptotically optimal effect, that is an equivalent increase of the trials.
	Finally, by optimizing tunable factor, we obtain an excellent non-Gaussian state, two-mode squeezed number state, that achieves optimal sensitivity and reaches the Heisenberg limit.

	\section*{Acknowledgments} 
	This work is supported by the National Natural Science Foundation of China (Grant No. 61701139).

\section*{Appendix} 
The details of two-mode squeezed vacuum in twin Fock representation can be found in a great of quantum optics textbooks. 
Here we focus on the expression of two-mode squeezed number state.
Let us start with number state $\left| {1,1} \right\rangle$ satisfying 
\begin{equation}
{\hat a^2}\left| {1,1} \right\rangle  = 0
\end{equation}
and two-mode Bogoliubov transformation
\begin{equation}
\hat S\left( \xi  \right)\hat a{\hat S^\dag }\left( \xi  \right) = \hat a\cosh r + {\hat b^\dag }\sinh r.
\end{equation}
Multiplying by squeezing operator $\hat S\left( \xi  \right)$ from the left and exploiting the unitary property of it, we can derive
\begin{eqnarray}
\nonumber \hat S\left( \xi  \right){{\hat a}^2}\left| {1,1} \right\rangle  &&= \hat S\left( \xi  \right)\hat a{{\hat S}^\dag }\left( \xi  \right)\hat S\left( \xi  \right)\hat a{{\hat S}^\dag }\left( \xi  \right)\hat S\left( \xi  \right)\left| {1,1} \right\rangle  \\ 
 \nonumber &&= {\left( {\hat a\cosh r + {{\hat b}^\dag }\sinh r} \right)^2}\hat S\left( \xi  \right)\left| {1,1} \right\rangle  \\ 
  &&= 0. 
\end{eqnarray}
We now decompose the squeezed number state into the Fock state
\begin{equation}
\hat S\left( \xi  \right)\left| {1,1} \right\rangle  = \sum\limits_{j,k}^\infty  {{C_{j,k}}\left| {j,k} \right\rangle} 
\end{equation}
in order to examine the photon statistics.
On the basis of the principle of the creation and the annihilation operators, and we only consider the solution which includes state $\left| {1,1} \right\rangle$, hence the solution is given by
\begin{equation}
{\sinh ^2}r \cdot {C_{j - 2,k - 2}} + \sinh \left(2r\right) \cdot {C_{j - 1,k + 1}} + {\cosh ^2}r \cdot {C_{j,k}} = 0.
\end{equation}
Using the property of progression we can supply the recurrence relation
\begin{equation}
{C_{j,k}} = {\left( { - \frac{{\sinh r}}{{\cosh r}}} \right)^{j - 1}}\left[ {j{C_{1,1}} + \frac{{\sinh r}}{{\cosh r}}\left( {j - 1} \right){C_{0,0}}} \right]{\delta _{j,k}}.
\end{equation}
Where $C_{0,0}$ and $C_{1,1}$ have been expressed in Eqs. (\ref{00}) and (\ref{11}), further, the expression of two-mode squeezed number state can be written as
\begin{equation}
{\left| \psi  \right\rangle _\textrm{SN}} = \sum\limits_{k = 0}^\infty  {{{\left( { - \tanh r} \right)}^{k - 1}}\left[ {k{C_{1,1}} + \tanh r\left( {k - 1} \right){C_{0,0}}} \right]\left| {k,k} \right\rangle }. 
\end{equation}
With the help of Fock state representation of two-mode squeezed vacuum, we can obtain the Eq. (\ref{TSB}).


%

\end{document}